\documentclass[traditabstract]{aa}
\usepackage{graphicx}
\usepackage{txfonts}
\usepackage{longtable}
\usepackage[authoryear]{natbib}
\bibpunct{(}{)}{;}{a}{}{,}

\def\ltsima{$\; \buildrel < \over \sim \;$}
\def\simlt{\lower.5ex\hbox{\ltsima}}
\def\gtsima{$\; \buildrel > \over \sim \;$}
\def\simgt{\lower.5ex\hbox{\gtsima}}

\begin{document}
   \title{The chemical composition of the stellar cluster {\sl Gaia1}: \\ no surprise behind Sirius}

   \author{A. Mucciarelli\inst{1,2}, L. Monaco\inst{3}, P. Bonifacio\inst{4}, I. Saviane\inst{5}}
         
      \offprints{A. Mucciarelli}

   \institute{Dipartimento di Fisica e Astronomia, Universit\`a degli Studi di Bologna, 
               Viale Berti Pichat, 6/2, I-40127 Bologna, Italy;
             \email{alessio.mucciarelli2@unibo.it} 
             \and
  			INAF - Osservatorio Astronomico di Bologna,
              Via Ranzani 1, 40127 Bologna, Italy
              \and
             Departamento de Ciencias Fisicas, Universidad Andres Bello, Fernandez Concha 700, Las Condes, Santiago, Chile
	     \and
	     GEPI, Observatoire de Paris, PSL Research University, CNRS, Place Jule Janssen 92190, 
	     Meudon, France
	     \and
	     European Southern Observatory, Alonso de Cordova 4860, Macul, Santiago, Chile
	     }

     \authorrunning{A. Mucciarelli et al.}
   \titlerunning{Gaia1}

   \date{Submitted to A\&A }

\abstract{We observed 6 He-clump stars of the intermediate-age stellar cluster Gaia1 
with the MIKE/MAGELLAN spectrograph.
A possible extra-galactic origin of this cluster, recently discovered thanks to the first 
data release of the ESA Gaia mission, has been suggested,
based on its orbital parameters. Abundances for Fe, $\alpha$, proton- and neutron-capture
elements have been obtained. We find no evidence of intrinsic abundance spreads.
The iron abundance is solar ([FeI/H]=+0.00 $\pm 0.01$; $\sigma = 0.03$ dex).
All the other abundance ratios are, by and large, solar-scaled,
similar to the Galactic thin disk and open clusters stars of similar metallicity. The
chemical composition of Gaia1 does not support an extra-galactic origin for this stellar 
cluster, that can be considered as a standard Galactic open cluster.}


\maketitle
%

\section{Introduction}
\label{intro}

{\sl Gaia1} is a stellar cluster that has been recently identified by \citet{koposov17} 
using the first data release of the ESA Gaia mission \citep{brown16}.
Its identification has been precluded for decades by 
its proximity ($\sim10$ arcmin) to the bright star Sirius.

Recently, \citet[][hereafter S17]{simpson17} performed a prompt spectroscopic follow-up of this system, 
confirming that {\sl Gaia1} is a stellar cluster, according to its
radial velocity (RVs) and [Fe/H].
They identified 41 cluster members, 27 of them observed with the high-resolution spectrograph HERMES and 
the other ones with the low-resolution spectrograph AAOmega, both at the Anglo-Australian Telescope.
The mean radial velocity is +58.30$\pm$0.22 km/s, with a dispersion of 0.94$\pm$0.15 km/s, 
while the mean iron abundance (from HERMES targets only) is [Fe/H]=--0.13$\pm$0.13 dex  
and its age is $\sim$3 Gyr.
Using Gaia and 2MASS positions for the cluster stars, they derived 
a first estimate of its proper motion and of its orbit, in particular
finding a Galactocentric distance ($R_{GC}$) of 11.8$\pm$0.2 kpc, a maximum height above the Galactic plane of 
$z_{max}=1.7^{+2.1}_{-0.9}$ kpc and an eccentricity $\epsilon=0.3\pm0.2$. 
These orbital parameters, taken at face values, could suggest 
an extra-galactic origin for {\sl Gaia1}, even if they have large 
uncertainties that prevent any firm conclusion.

In this paper we present chemical abundances of Fe, Na, Mg, Al, Si, Ca, Ti, Ba and Eu 
for 6 giant stars observed at the Magellan II Telescope.


\section{Observations}
\label{obs}
\begin{figure}
\centering
\includegraphics[width=\columnwidth]{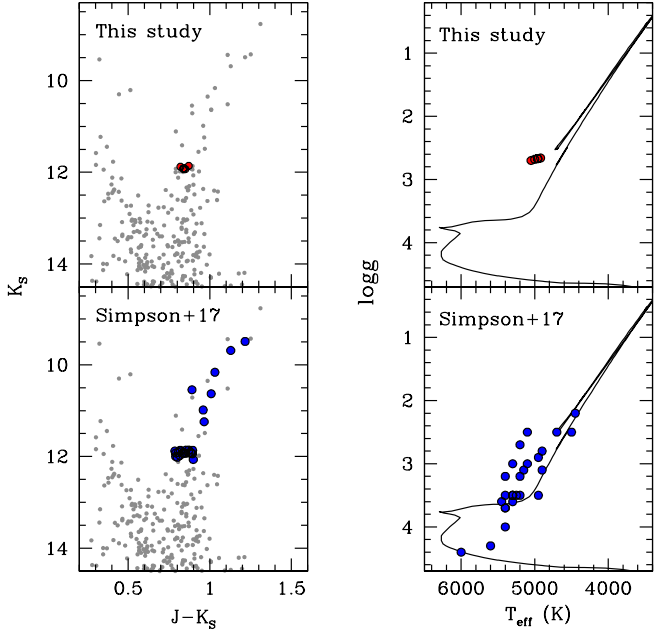}
  \caption{Left panels: the 2MASS ($K_{S}$ vs (J-$K_{S}$)) color-magnitude diagram 
     of {\sl Gaia1}  (grey points) with marked the position of our targets (red points) and 
     of those by S17 (blue points).
     Right panels: position of the MIKE and HERMES targets
     in the $T_{eff}$--log~g plane, superimposed to a theoretical isochrone by \citet{bressan12} 
     with [Fe/H]=~0.0 and age 3 Gyr. Unexpectedly, the S17 stars do not define an RGB in the 
     theoretical plane, suggesting that their parameters are not correct.}
     \label{cmd}
 \end{figure}

Spectra of 7 stars having infrared colors and magnitudes 
compatible with being He-clump stars of the stellar cluster {\sl Gaia1} have been secured 
using the Magellan Inamori Kyocera Echelle (MIKE) spectrograph 
\citep{bernstein03} mounted on the Magellan~II Telescope at Las Campanas 
Observatory. 
Table~\ref{info} lists the coordinates, the 2MASS magnitudes, RVs and atmospheric 
parameters for the observed targets.
The position of the observed targets in the 2MASS ($K_{S}$ vs (J-$K_{S}$)) color-magnitude diagram 
is shown in left-upper panel of Fig.~\ref{cmd} as red points. Left-lower panel shows (as blue points) 
the position of the targets observed with HERMES by S17. 
All the MIKE targets have been observed with the 0.7" x 5.0" slit, providing 
a spectral resolution of $\sim$36000 in the red arm (covering from $\sim$5000 to 
$\sim$9150 \AA\ ).
Exposure times varied from 600 to 1800 sec. 
The signal-to-noise ratio ranges from $\sim$40 to $\sim$80 at $\sim$6000 \AA\ .
Bias-subtraction, flat-fielding, spectral extraction and wavelength 
calibration have been performed using the CarPy MIKE pipeline \citep{kelson03}.

RVs have been measured through {\tt DAOSPEC} \citep{stetson08} 
by measuring the position of almost 400 metallic and unblended lines selected in the red spectrum.
Six among the target stars share very similar RVs (see Table~\ref{info}) and are classified as members of Gaia1. 
Star \#3, instead, presents a different RV and damped absorption lines, and is, as such, considered not member and will not 
be further discussed.
The mean heliocentric RV is +57.6$\pm$0.4 km/s ($\sigma$=~1.0 km/s) 
that nicely matches with the value derived by S17.

\begin{table*}
  \begin{center}
  \caption{Main information on the observed MIKE targets: 
  2mass identification number, 2MASS J and $K_{S}$ magnitudes, RV and atmospheric parameters 
  derived from the MIKE spectra, together with the SNR measured at $\sim$600nm. }
  \label{info}
  \tiny
  \begin{tabular}{lcccccccccc}
  \hline
     ID &  ID$_{2MASS}$ & RA$_{J2000}$  &Dec$_{J2000}$  & J    &  $K_{S}$  &  RV  &  T$_{eff}$   &log g   & v$_t$   &  SNR   \\
     
        &             &  [deg]        & [deg]	      &      &     & {\tiny [km/s]}  &[$^o$K] &{\tiny [cm s$^{-2}$]}& {\tiny [km/s]} &  @600nm \\
\hline
   1 &   2MASS06455819-1641596     &	101.492499 & -16.699909      & 12.731  & 11.860 & +58.2&  5000$\pm$60  &   2.68$\pm$0.10  & 1.5$\pm$0.1   &  60    \\ 
   2 &   2MASS06454837-1643113     &	101.451559 & -16.719822      & 12.782  & 11.932 & +58.4&  4960$\pm$80  &   2.67$\pm$0.12  & 1.5$\pm$0.1   &  40    \\ 
   3 &   2MASS06454801-1642240     &    101.450069 & -16.706678      & 12.832  & 11.996 & +64.5&  ---          &   ---           & ---	         &  40	 \\
   4 &   2MASS06455237-1643471     &	101.468234 & -16.729773      & 12.698  & 11.878 & +57.2&  5050$\pm$80  &   2.70$\pm$0.12  & 1.4$\pm$0.1   &  40     \\
   5 &   2MASS06455379-1645521     &	101.474151 & -16.764490      & 12.770  & 11.930 & +57.4&  4960$\pm$60  &   2.67$\pm$0.10  & 1.4$\pm$0.1   &  70    \\ 
   6 &   2MASS06460723-1647294     &	101.530159 & -16.791500      & 12.753  & 11.913 & +56.0&  4920$\pm$50  &   2.66$\pm$0.10  & 1.3$\pm$0.1   &  80    \\ 
   7 &   2MASS06455253-1640582     &	101.468910 & -16.682854      & 12.758  & 11.918 & +58.5&  4960$\pm$50  &   2.67$\pm$0.10  & 1.6$\pm$0.1   &  80    \\ 
\hline
\end{tabular} 
\end{center}
\end{table*}


\section{Chemical analysis}
\label{data}

\subsection{Atmospheric parameters}
The atmospheric parameters have been derived as follows: 
{\sl (i)}~effective temperatures ($T_{eff}$) have been derived 
spectroscopically from the excitation equilibrium thanks 
to the large number ($\sim$120) of measured Fe~I lines;
{\sl (ii)}~surface gravities (log~g) have been obtained from the Stefan-Boltzmann 
relation, adopting the spectroscopic $T_{eff}$, the bolometric corrections 
calculated according to \citet{buzzoni10}, the true distance modulus 
$(m-M)_0$=~13.3 (S17 from the 2MASS photometry) and the 
color excess E(B-V)=~0.41 mag obtained from the maps by \citet{schlegel98} corrected according to 
\citet{bonifacio00}. For all the stars a stellar mass of 1.5 $M_{\odot}$ has been 
adopted;
{\sl (iii)}~microturbulent velocities (v$_t$) have been obtained by minimizing the trend 
between iron abundance and line strength.

The advantage of this {\sl hybrid} approach is to exploit at the best 
all the information in hand, both from spectroscopy and photometry, 
and to minimize the uncertainties in the color excess that affect mainly  
the photometric $T_{eff}$\footnote{Note that the uncertainty in the color excess 
marginally impacts in the determination of log~g: 
values of E(B-V) in the direction of {\sl Gaia1} range from 0.36 to 0.66 mag 
(see S17 and references therein).
A variation of $\pm$0.1 mag in E(B-V) leads to a variation in log~g 
of $\pm$0.002, but has a significant ($\pm$140 K) impact on $T_{eff}$ even when they are 
derived from a color marginally affected by the reddening as $(J-K_{S})$.}, 
as well as possible systematics in the ionization equilibrium 
due to over-ionization effects that affect the spectroscopic log~g.


Right-upper panel of Fig.~\ref{cmd} shows the position of the observed targets 
in the $T_{eff}$-log~g plane (red points) in comparison with a theoretical isochrone 
from Padua database \citep{bressan12} with solar metallicity and 
an age of 3 Gyr (see S17). We find that within the uncertainties the atmospheric parameters 
(derived as explained above) reasonably fit the position expected for the He-clump.
On the other hand, an inspection on the atmospheric parameters derived by 
S17 for their targets reveals a significant discrepancy with 
the atmospheric parameters expected for their evolutionary stage 
(right-lower panel of Fig.~\ref{cmd}). 
Even if their observed targets belong to the He-clump or to the bright red giant branch, 
 $T_{eff}$ and log~g are compatible with those of less evolved 
stars. In particular, half of their sample (14 out 27 stars) have log~g higher than 3.5, 
that are values unlikely for the observed targets and a systematic offset toward higher 
$T_{eff}$ seems to be present, with the extreme case of a He-clump star for which they estimated
$T_{eff}$=~6000 K and log~g=~4.4. 
Note that S17 derived all the atmospheric parameters spectroscopically: 
we stress that this approach can be very dangerous in case of low-quality spectra and/or 
when the number of Fe lines are not large enough to guarantee a robust coverage in excitation 
potential and line strength, or when the Fe~II lines are too few/too weak.

\subsection{Sanity check}
As a sanity check we derived the atmospheric parameters using different methods.
First, $T_{eff}$ and log~g have been derived from the photometry, 
adopting the $(J-K)_0$-$T_{eff}$ transformation by \citet{ghb}. 
The photometric $T_{eff}$ are on average lower than spectroscopic ones by 
$\sim$200-300 K. However, a significant (at a level of 3-5 $\sigma$) 
positive slope between iron abundances and excitation potential is found for all the targets, 
pointing out that the photometric $T_{eff}$ are not totally correct, probably 
due to the large uncertainties in the color excess. The use of 
photometric parameters leads to a decrease of [Fe/H] of about 0.2 dex.
On the other hand, a fully spectroscopic determination of all the parameters 
provides $T_{eff}$ very similar to that obtained above and log~g higher by $\sim$0.3, 
but with a negligible impact on the abundance ratios discussed here.

\subsection{Abundance determination}
Abundances for Fe, Na, Al, Si, Ca and Ti have been derived 
from the equivalent widths of unblended transitions (measured with the code {\tt DAOSPEC}) 
and using the code {\tt GALA} \citep{mu13} based on the suite of 
software developed by R. L. Kurucz\footnote{http://kurucz.harvard.edu/programs.html, http://wwwuser.oats.inaf.it/castelli/sources.html}.
Abundances for Mg, Ba and Eu  have been derived from spectral synthesis, 
because the used Mg lines ($\sim$6318-19 \AA\ ) are located on the red wing 
of an auto-ionization Ca line, while the Ba and Eu lines are affected by 
isotopic and hyperfine splittings. 
Solar reference abundances are from \citet{gs98}. 
The linelist and the determination of the abundances uncertainties are described 
in Appendix A and B, respectively.


\begin{table*}
  \begin{center}
  \caption{Measured chemical abundance ratios for individual MIKE targets.
  Last row lists the mean abundances and the corresponding 1$\sigma$ uncertainty.}
  \label{abu}
  \tiny
  \begin{tabular}{lccccccccc}
  \hline
     ID &  [FeI/H]        &   [Na/Fe]   &  [Mg/Fe]   &  [Al/Fe]   &   [Si/Fe]   &   [Ca/Fe]   &  [Ti/Fe]   &  [BaII/Fe]   &   [EuII/Fe]   \\
   \hline
   1 &   -0.02$\pm$0.07   &  --0.14$\pm$0.05  &  +0.02$\pm$0.04  &  +0.13$\pm$0.05  & +0.01$\pm$0.07  & --0.03$\pm$0.05 &  +0.02$\pm$0.06   &  +0.26$\pm$0.10  &--0.04$\pm$0.06     \\
   2 &   +0.05$\pm$0.06   &  --0.02$\pm$0.07  & --0.02$\pm$0.07  &  +0.01$\pm$0.10  & +0.03$\pm$0.08  & --0.06$\pm$0.06 & --0.01$\pm$0.06   &  +0.14$\pm$0.08  & +0.00$\pm$0.08      \\
   4 &   -0.02$\pm$0.07   &  --0.13$\pm$0.10  &  +0.09$\pm$0.07  &  +0.08$\pm$0.07  &--0.02$\pm$0.08  & --0.08$\pm$0.03 &  +0.00$\pm$0.05   &  +0.21$\pm$0.05  &--0.06$\pm$0.08      \\
   5 &   +0.00$\pm$0.06   &  --0.08$\pm$0.05  &  +0.00$\pm$0.05  &  +0.13$\pm$0.04  &--0.01$\pm$0.07  & --0.09$\pm$0.04 & --0.04$\pm$0.07   &  +0.23$\pm$0.10  & +0.15$\pm$0.06      \\
   6 &   +0.01$\pm$0.06   &  --0.08$\pm$0.06  &  +0.06$\pm$0.07  &  +0.07$\pm$0.05  & +0.03$\pm$0.08  & --0.04$\pm$0.07 & --0.04$\pm$0.07   &  +0.22$\pm$0.11  & +0.10$\pm$0.07      \\
   7 &   -0.03$\pm$0.06   &  --0.02$\pm$0.06  &  +0.11$\pm$0.07  &  +0.11$\pm$0.04  &--0.03$\pm$0.07  & --0.01$\pm$0.05 &  +0.02$\pm$0.06   &  +0.15$\pm$0.11  & +0.07$\pm$0.07      \\
  \hline
   Mean ($\sigma$) &  +0.00 (0.03) &  --0.08 (0.05)  &  +0.04 (0.05)  &  +0.09 (0.04)  & +0.00 (0.03)  &  
  --0.05 (0.03)  & --0.01 (0.03)  &  +0.20 (0.05) &  +0.04 (0.08) \\
\hline
\end{tabular} 
\end{center}
\end{table*}


\section{The chemical composition of {\sl Gaia1}}

\subsection{Iron content}

{\sl Gaia1} has an average iron abundance of [FeI/H]=+0.00$\pm$0.01 dex ($\sigma$=~0.03 dex).
We used the maximum likelihood (ML) algorithm described in \citet{mu12} 
to estimate whether the observed scatter is compatible or not 
with a null intrinsic spread, taking into account the uncertainties 
of individual stars. The observed [FeI/H] spread turns out to be 
fully consistent with a null intrinsic spread.
The same conclusions are obtained adopting the pure spectroscopic and 
photometric sets of atmospheric parameters. The former set 
of parameters provides a very similar abundance,
while the latter one provides an average abundance lower by $\sim$0.2 dex because of the 
lower $T_{eff}$.

S17 derived from 27 giant stars observed with HERMES 
an average abundance [Fe/H]=--0.13$\pm$0.03 dex ($\sigma$=~0.13 dex). 
Using their quoted uncertainties, the ML algorithm suggests the presence 
of a non-null intrinsic spread, $\sigma_{int}$=~0.12$\pm$0.01 dex, 
at variance to our analysis. 
Even if their sample is significantly larger than ours, 
the intrinsic iron spread obtained from their abundances is likely 
due to two effects: (1)~the large uncertainty in their spectroscopic parameters 
(see Fig.~\ref{cmd}), and (2)~the fact that their [Fe/H]
uncertainties do not include the contribution of the atmospherical parameters errors, 
hence under-estimating the total errorbar.
Even if the metallicities derived by S17 and in this study are 
consistent, we discourage the use of their atmospheric parameters that 
turn out to be inconsistent with the evolutionary stages of the observed stars.

\subsection{$\alpha$-elements}

\begin{figure}
\centering
\includegraphics[width=\columnwidth]{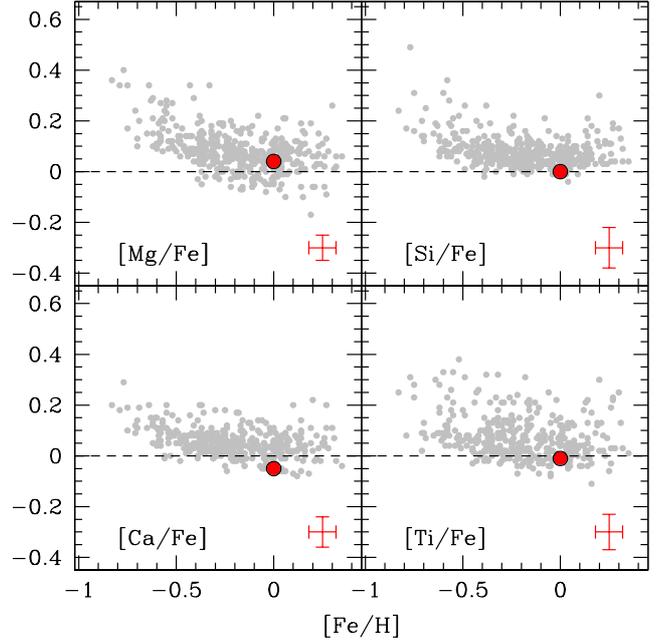}
  \caption{The behaviour of the average [Mg/Fe], [Si/Fe], [Ca/Fe] and [Ti/Fe] 
     abundance ratios of {\sl Gaia1} as a function of [Fe/H] (red point), 
     in comparison with Galactic thin disk stars \citep[grey points,][]{soubiran05}}.
     \label{alfa}
 \end{figure}

The measured [$\alpha$/Fe] abundance ratios (see Table~2) turn out to be 
solar-scaled, 
pointing out that the cluster formed from a gas enriched both from 
Type II and Type Ia supernovae. 
Fig.~\ref{alfa} shows the behaviour of the measured [$\alpha$/Fe] abundance ratios 
as a function of [Fe/H] in comparison with the Galactic thin disk stars 
\citep[grey points,][]{soubiran05}. 
For all the abundance ratios, {\sl Gaia1} well matches the mean locus 
described by the Galactic field stars, suggesting a strong similarity with the 
Galactic thin disk. 

\subsection{Na and Al}

Na and Al are usually associated to nucleosynthesis by proton-captures. 
We derived for {\sl Gaia1} [Na/Fe]=--0.08$\pm$0.02 ($\sigma$=0.05 dex)  and [Al/Fe]=+0.09$\pm$0.02 ($\sigma$=0.05 dex).
As visible in the upper panels of Fig.~\ref{other}, at the same metallicity of the cluster 
the thin disk stars are essentially solar-scaled and {\sl Gaia1} well matches, 
for both the abundance ratios, the observed Galactic trend.
Note that [Na/Fe] values shown in Fig.~\ref{other} for Galactic stars do not include 
corrections for non-local thermodynamical equilibrium. In order to provide a homogeneous 
comparison with the literature data for the thin disk stars, we did not taken into account 
such corrections in our measured [Na/Fe]. However, corrections for the targets estimated 
according to \citet{lind11} are of about --0.1 dex and they do not change significantly 
our conclusions.

No evidence of intrinsic Na and Al star-to-star scatters is found for these stars. 
Such chemical inhomogeneities are commonly observed among globular cluster stars 
\citep[see][and references therein]{gratton12} and usually explained within a framework 
of a self-enrichment process. On the other hand, open cluster stars, in virtue of their 
lower mass and density, do not undergo to self-enrichment processes and they show 
homogeneous Na and Al contents. The lack of significant star-to-star variations in {\sl Gaia1} 
(despite the small number of observed stars) agrees with its current low mass 
\citep[$\sim10^4 M_{\odot}$][]{,koposov17}. 
It is also likely that the cluster suffered 
only limited mass loss in the $\sim$3 Gyr that elapsed since its formation.

\subsection{Neutron-capture elements: Ba and Eu}
\begin{figure}
\centering
\includegraphics[width=\columnwidth]{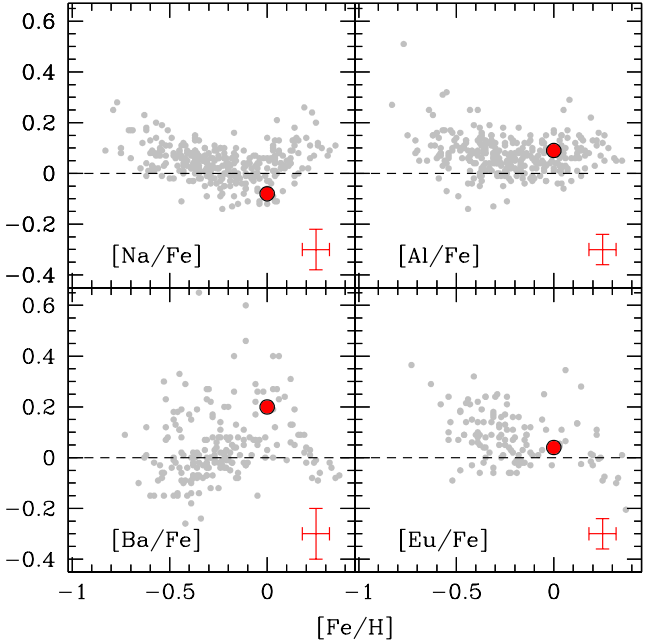}
  \caption{The behaviour of the average [Na/Fe], [Al/Fe], [Ba/Fe] and [Eu/Fe] 
     abundance ratios of {\sl Gaia1} as a function of [Fe/H] (red point), 
     in comparison with  Galatic thin disk stars  
     (grey points, \citet{soubiran05} for Na and Al, and 
     \citet{reddy03,bensby05} for Ba and Eu)}.
     \label{other}
 \end{figure}

We measured Ba, as prototype of elements produced through slow neutron captures, 
and Eu, mainly produced through rapid neutron captures, obtaining 
[Ba/Fe]=+0.20$\pm$0.02 ($\sigma$=~0.05 dex) and 
[Eu/Fe]=+0.04$\pm$0.03 ($\sigma$=~0.08 dex), respectively.
Lower-panels of Fig.~\ref{other} show the behaviour of [Ba/Fe] and [Eu/Fe] 
abundance ratios measured in {\sl Gaia1}: also for these two elements 
we find a good agreement with the Galactic thin disk stars of similar metallicity.

\section{Conclusions}

The chemical composition of {\sl Gaia1} that we derived from MIKE spectra 
well matches with that of thin disk stars and open clusters 
with similar metallicity \citep[see e.g.][and reference therein]{pancino10,mish15}. This nice match has been found for all the main 
groups of elements, i.e. $\alpha$, proton- and neutron-capture elements, 
indicating that this cluster formed from a gas that has had a chemical enrichment 
similar to that of the Galactic thin disk. 
A possible extra-galactic origin of {\sl Gaia1} 
is not supported by the comparison between its chemical composition and that of 
other stellar systems.
The galaxies currently populating the Local Group are more metal-poor than 
{\sl Gaia1} and they do not reach solar metallicity 
\citep{mcconnachie12}. The only extra-galactic environment approaching similar metallicities 
is the Sagittarius dwarf spheroidal galaxy but its most metal-rich stars 
are characterized by sub-solar [$\alpha$/Fe] abundance ratios and enhanced [Ba/Fe] and [Eu/Fe] 
\citep{bonifacio00sgr,monaco05,sbordone07}.

\begin{figure}
\centering
\includegraphics[width=\columnwidth]{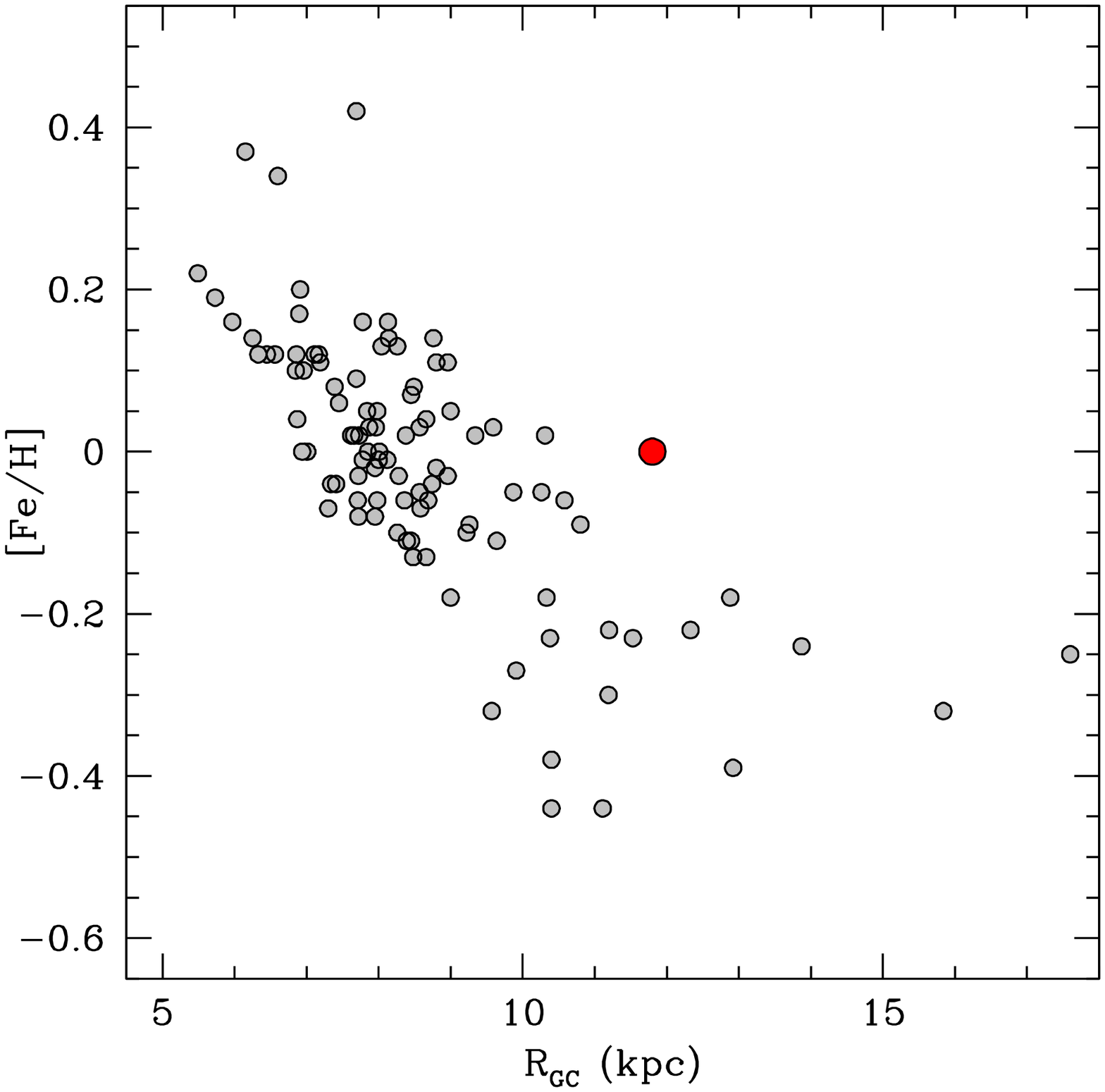}
  \caption{ The behaviour of [Fe/H] as a function of the Galactocentric distance 
  for open clusters \citep[][grey circles]{netopil16,magrini17} and 
  {\sl Gaia1} (red point).}
     \label{trend}
 \end{figure}

The mean metallicity of the thin disk (both field and open cluster stars) 
is known to decrease at large galactocentric distances
\citep{pancino10,hayden15,netopil16,magrini17}.
As shown in Fig.~\ref{trend},  
at the distance of {\sl Gaia1}, the mean metallicity of open clusters is 
about --0.3 dex lower than that of the cluster, even if a large 
dispersion is observed at these distances.
Solar metallicity open clusters are typically found at $R_{GC}$ 
of $\sim$7-8 kpc. 
On the other hand, stars with solar metallicity are present in the thin disk 
also at similar distances: the metallicity distribution for thin disk stars 
provided by \citet{hayden15} at 11$<R_{GC}<$13 kpc and 1$<\mid{z}\mid<$2 kpc 
is peaked at [Fe/H]=--0.38 dex but reaching super-solar metallicities.

Our results on the chemical composition strongly argue against an extra-galactic origin 
for {\sl Gaia1}. 
In fact it appears to be an unremarkable standard Galactic open cluster. 
Its position with respect to the overall 
trend between [Fe/H] and $R_{GC}$ could suggest that it formed in the 
inner disk, progressively migrating toward higher $R_{GC}$, thus 
explaining its possible peculiar orbit with respect to other open clusters. 
However, it is worth noticing that the precise value of $R_{GC}$ can be affected 
by the value of E(B-V) that remains still uncertain for this cluster.
We checked the measured abundance ratios with those derived by 
\citet{nissen16} for solar twin stars of ages 
similar to that of {\sl Gaia1} in the solar neighborhood. 
Within the uncertainties the abundances of {\sl Gaia1} are comparable 
with those of the solar neighborhood, but for [Al/Fe] and [Ba/Fe] that in this cluster 
are higher (by $\sim$0.1 dex) with respect to coeval solar twin stars. 
However, we cannot totally rule out that {\sl Gaia1} originally formed 
in the solar neighborhood.

Although the orbital parameters
inferred by S17 may suggest that the cluster has been accreted by
the Milky Way, they are still, within uncertainties, fully compatible
with the majority of known Galactic open clusters.
The uncertainty on the orbit of Gaia1 is dominated by
the uncertainty on its proper motion. The situation will be
greatly improved with the second Gaia data release, and we defer
any conclusion on its kinematics to that time.

\begin{acknowledgements}
We thanks the anonymous referee for the useful comments and suggestions.
We are grateful to P. Di Matteo for useful
discussions on the Galactic metallicity gradient.
LM acknowledges support from "Proyecto interno" of the Universidad Andres Bello.
PB acknowledges financial support from the Scientific Council of Observatoire de Paris and from the action
f\'ed\'eratrice  ``Exploitation Gaia''.
\end{acknowledgements}

\bibliographystyle{apj}

\clearpage
\begin{appendix}

\section{Linelist}

The analysed lines have been selected according to a synthetic spectrum 
calculated with the code {\tt SYNTHE}, adopting a model atmosphere calculated 
with {\tt ATLAS9} with the representative atmospheric parameters of the stars 
(that have $T_{eff}$, log~g and $v_t$ very similar each other). 
The reference synthetic spectrum has been computed including all the atomic and molecular 
transitions of the last version of the Kurucz/Castelli 
linelists\footnote{http://wwwuser.oats.inaf.it/castelli/linelists.html}, updating the 
oscillator strengths for some transitions of interest (as explained below). 
For the abundances calculated using the equivalent widths (EW) we selected only lines predicted to be unblended.
Atomic data for Fe~I lines are from \citet{martin88} and \citet{fuhr06}, while 
those for Fe~II lines from \citet{melendez}. Oscillator strengths for Ca~I lines 
are mainly from \citet{sr81}, for Ti~I lines are from \citet{martin88} and \citet{lawler13}.
For Si~I lines we adopted, when available, the furnace oscillator strengths by 
\citet{garz}, while for other lines, for which the available log~gf have large uncertainties 
or badly reproduce the solar spectrum, we derived astrophysical oscillator strengths 
using the solar flux spectra of \citet{neckel} and a solar model atmosphere 
calculated with the chemical mixture of \citet{gs98}. The same method has been adopted 
to compute the oscillator strengths for the Al~I doublet at 6696-6698 \AA\ because the 
available values in literature underestimate the solar abundance of about 0.2 dex.
For the two used Na~I doublets at 5682-88 \AA\ and 6154-6160 \AA\ we employed the 
log~gf available in the NIST database\footnote{http://physics.nist.gov/PhysRefData/ASD/lines\_form.html}, 
as well as for the Mg~I doublet at 6318-6319 \AA\ .
For the transitions measured using spectral synthesis because affected by hyperfine and 
isotopic splitting, we adopted the linelists available in the Kurucz/Castelli database 
(Ba~II lines) and that provided by \citet{lawler01} (Eu~II line).
Table A.1 lists for all the used lines the measured EW and the adopted oscillator strength
and excitation potential.

\begin{table}[h]
  \begin{center}
  \caption{Wavelength, oscillator-strength, excitation potential and measured EW 
  for the used transitions. The complete version of the table is available in electronic form.}
  \label{info}
  \tiny
  \begin{tabular}{lccccc}
  \hline
     ID &  $\lambda$ & Element  & log~gf  & $\chi$   &  EW  \\
        &   (\AA\ )  &   &   &  (eV)  & (m\AA\ ) \\
\hline

   1 & 5197.9	& 26.00  &   -1.620  &   4.300   &   45.30  \\ 
   1 & 5236.2	& 26.00  &   -1.497  &   4.190   &   71.10  \\ 
   1 & 5249.1	& 26.00  &   -1.460  &   4.470   &   47.60  \\ 
   1 & 5315.1	& 26.00  &   -1.550  &   4.370   &   55.10  \\ 
   1 & 5326.8	& 26.00  &   -2.100  &   4.410   &   25.60  \\ 
   1 & 5379.6	& 26.00  &   -1.514  &   3.690   &   83.20  \\ 
   1 & 5386.3	& 26.00  &   -1.740  &   4.150   &   45.90  \\ 
   1 & 5398.3	& 26.00  &   -0.710  &   4.450   &   96.60  \\ 
   1 & 5405.4	& 26.00  &   -1.390  &   4.390   &   69.60  \\ 
   1 & 5412.8	& 26.00  &   -1.716  &   4.430   &   40.00  \\ 
   1 & 5470.1	& 26.00  &   -1.790  &   4.450   &   33.30  \\ 
   1 & 5476.3	& 26.00  &   -0.935  &   4.140   &  100.60  \\ 
   1 & 5491.8	& 26.00  &   -2.188  &   4.190   &   34.40  \\ 
   1 & 5522.4	& 26.00  &   -1.520  &   4.210   &   71.90  \\ 
   1 & 5525.5	& 26.00  &   -1.084  &   4.230   &   87.20  \\ 
   1 & 5543.9	& 26.00  &   -1.110  &   4.220   &   92.60  \\ 
   1 & 5577.0	& 26.00  &   -1.550  &   5.030   &   25.40  \\ 
   1 & 5609.0	& 26.00  &   -2.400  &   4.210   &   14.30  \\ 
   \hline
\end{tabular} 
\end{center}
\end{table}

\section{Uncertainties}

The total uncertainty in the measured [X/Y] abundance ratio has been 
computed by adding in quadrature two terms:
\begin{enumerate}
\item the error related to the line measurement. 
For those elements for which the abundance has been derived through the measured EWs, 
this term has been estimated as the dispersion of the mean divided by the root mean 
square of the number of lines.
For Mg, Ba and Eu, for which we used spectral synthesis, this error has been estimated 
using Montecarlo simulations, injecting Poissonian noise into the best-fit spectrum in order 
to reproduced the measured SNR and creating for each line a sample of 500 noisy synthetic spectra. 
These spectra have been re-analysed with the same procedure used for the observed lines and 
the 1$\sigma$ dispersion of the abundance distribution taken as uncertainty. 
\item the error related to the atmospheric parameters. This uncertainty 
has been calculated by varying each time only one parameter by the corresponding error, 
keeping the ones fixed and repeating the analysis. 
Uncertainties in spectroscopic $T_{eff}$ and $v_t$ have been estimated according the error 
in the slope between excitation potential and iron abundances, and between 
reduced EW ($\log(EW/\lambda)$) and iron abundances, respectively \citep[see][for details]{mu13}. 
Uncertainties in surface gravities have been computed by including the 
errors in spectroscopic $T_{eff}$, mass, reddening and distance modulus.
\end{enumerate}

\end{appendix}

\end{document}